\documentclass[epj]{svjour}
\usepackage[colorlinks,citecolor=blue,linkcolor=blue,anchorcolor=blue,filecolor=blue,
urlcolor=blue]{hyperref}
\usepackage{graphicx}
\usepackage{color}
\usepackage[dvipsnames]{xcolor}
\colorlet{orange}{orange!120!}
\usepackage{ulem}
\usepackage{dcolumn}
\usepackage{bm}
\usepackage{amsmath}
\usepackage{amssymb}
\usepackage{amsfonts}
\usepackage{mathtools}
\usepackage{cite}

\begin{document}
\emergencystretch 3em	

\title{Do short range correlations inhibit the appearance of the nuclear pasta?}
\titlerunning{Do short range correlations inhibit the appearance of the nuclear pasta?}

\author{M. R. Pelicer$^1$ \and D. P. Menezes$^1$ \and M. Dutra$^{2,3}$ \and O. Louren\c{c}o$^{2,3}$}
\authorrunning{
M. R. Pelicer et al.
}

\institute{$^1$Depto de F\'{\i}sica - CFM - Universidade Federal de Santa Catarina  Florian\'opolis - SC - CP. 476 - CEP 88.040 - 900 - Brazil \\
\mbox{$^2$Departamento de F\'isica, Instituto Tecnol\'ogico de Aeron\'autica, DCTA, 12228-900, S\~ao Jos\'e dos Campos, SP, Brazil} \\
\mbox{$^3$Universit\'e de Lyon, Universit\'e Claude Bernard Lyon 1, CNRS/IN2P3, IP2I Lyon, UMR 5822, F-69622, Villeurbanne, France}
}

\abstract{
It is well known that strongly correlated neutron-proton pairs, the short-range correlations (SRC), can modify many of the nuclear properties.
In this work we have introduced, for the first time, short range correlations in the calculation of the nuclear pasta phase at zero temperature and checked how they affect its size and internal structure. We have used two different parameterizations of relativistic models in a mean field approximation and the coexistence phase approximation as a first estimation of the effects. We have seen that for very asymmetric neutron-proton-electon matter, the pasta phase shrinks considerably as compared with the results without SRC and all internal structures vanish, except the simple spherically symmetric one, the droplets. Our results indicate a possible disappearance of these complicated structures as the temperature increases. 
\PACS{
			XX.XX.XX No PACS code given
		} 
}
\maketitle

\section{Introduction}
Nonspherical complex structures that appear due to a frustration in subsaturation nuclear densities are believed to be present in the inner crust of neutron stars and in core-collapse supernova cores~\cite{PhysRevLett.50.2066,hashimoto84}. Although possibly present only in a small range of densities and temperatures, different studies suggest that these exotic structures may have considerable impact on different astrophysical phenomena and on the magnetic evolution of neutron stars~\cite{Pons2013}. Some of the consequences of the existence of the pasta phase are expected to leave signatures in quasiperiodic oscillations observed in soft-gamma ray repeaters~\cite{soft} and magnetar giant flares due to crust thermal relaxation due to neutrino trapping~\cite{Okamoto:2013tja,PhysRevC.83.035803, PhysRevC.70.065806, Lin:2020nxy}.
More recently, the detection of late-time neutrinos from a Milky Way core-collapse supernova seems to be approaching reality~\cite{latetime} and it was claimed that neutrino diffusion seems to be affected by the pasta phase in protoneutron stars in such a way that its signal can be greatly enhanced after core collapse~\cite{neutrino}. 

From the theoretical point of view, there are some models and simulations that predict that the pasta structures can be very complex indeed. Calculations departed from the more traditional  1D, 2D, and 3D geometries in a single unit cell to density fluctuations~\cite{flut1,Pelicer:2021ils} as well as to structures resembling waffle, parking garage and triply periodic minimal surfaces (TPMS)~\cite{complex1,complex2,complex3}. Different calculations also show that the pasta phase size decreases as temperature increases and may occupy just a small portion between two homogeneous phases at certain temperatures~\cite{oldpasta,PRC82,PRC85,PhysRevC.105.025806,PhysRevD.106.063020,PhysRevC.103.055810,PhysRevC.102.015806}.

Most of the pasta phase calculations depend on equations of state (EOS) parameterized to satisfy nuclear matter bulk properties. One  missing ingredient in all pasta calculations is the inclusion of short-range correlations (SRC). Strongly correlated neutron-proton pairs can modify the internal structure of the nucleus and generate a series of non-trivial consequences~\cite{sciencesrc1,naturesrc2,naturescr3,naturesrc4,hen2017,duer2019,rev3,cai,baoanli21,baoanli22,baoanli-aop,lucas,dmnosso1,dmnosso2,dmnosso3}. In a seminal paper on cold dense matter obtained with the inclusion of SRC~\cite{science-src}, the authors claimed that the difference between the types of pairs (proton-proton, neutron-neutron and proton-neutron) {\it is due to the nature of the strong force and has implications for understanding cold dense nuclear systems such as neutron stars}. Hence, the introduction of SRC in the pasta phase structure is long due. 

So...what if the complicated pasta structures are simply not there when SRC are taken into account? As a first study, we investigate the effects of the SRC on the pasta phase and its structure with a simple prescription, the coexistence phase approximation (CPA)~\cite{PhysRevC.72.015802,oldpasta} within relativistic mean field models at zero temperature. In fact, as already mentioned, different approaches have been used to estimate the size and structure of the pasta phase and the CPA is a simple one and, as shown in \cite{oldpasta}, it underestimates the pasta size as compared with the Thomas-Fermi approximation (TF), which in turn, is also very simple when compared with recent calculations that take into account fluctuations and different structure coexistence \cite{complex2,PhysRevC.105.025806,Pelicer:2021ils}. Hence, CPA is used as a departing point to evaluate SRC effects and more sophisticated treatments will be used in future calculations.

In this paper, we start by studying the effects of the SRC on the homogeneous phase with different parametrizations, namely IUFSU~\cite{IUFSU}, FSU2R~\cite{FSU2R} and NL3~\cite{Lalazissis:1996rd}. The latter is chosen due to its 
different behaviour, with a much higher slope of the symmetry energy.
When SRC are included, the momentum~($k$) distribution present in the kinetic terms of the energy density, pressure, and scalar density is modified by the inclusion of a ``high momentum tail'' proportional to $k^{-4}$ and, as a consequence, the model has to be reparameterized in order to reproduce nuclear bulk properties. As a second step, we calculate the pasta phase and obtain its internal structures for the two models mentioned above.
 . 

The most important equations of the formalism are presented in the next section and the results are shown and discussed afterwards.

\section{Formalism}

We next introduce the Lagrangian density and the expressions used to calculate the SRC. The CPA method utilized to construct the pasta phase is subsequently shown.

\subsection{Relativistic model with short-range correlations}

We closely follow the formalism developed in Refs.~\cite{cai,lucas,dmnosso1,dmnosso2,dmnosso3} to construct the main equations of state of the model with SRC included. Firstly, we present the Lagrangian density, given by
\begin{align}
&\mathcal{L} = \overline{\psi}(i\gamma^\mu\partial_\mu - M_N)\psi 
+ g_\sigma\sigma\overline{\psi}\psi 
- g_\omega\overline{\psi}\gamma^\mu\omega_\mu\psi
\nonumber \\ 
&- \frac{g_\rho}{2}\overline{\psi}\gamma^\mu\vec{\rho}_\mu\vec{\tau}\psi
+\frac{1}{2}(\partial^\mu \sigma \partial_\mu \sigma - m^2_\sigma\sigma^2)
- \frac{A}{3}\sigma^3 - \frac{B}{4}\sigma^4 
\nonumber\\
&-\frac{1}{4}F^{\mu\nu}F_{\mu\nu} 
+ \frac{1}{2}m^2_\omega\omega_\mu\omega^\mu 
+ \frac{C}{4}(g_\omega^2\omega_\mu\omega^\mu)^2 -\frac{1}{4}\vec{B}^{\mu\nu}\vec{B}_{\mu\nu} 
\nonumber \\
&+ \frac{1}{2}\alpha'_3g_\omega^2 g_\rho^2\omega_\mu\omega^\mu
\vec{\rho}_\mu\vec{\rho}^\mu + \frac{1}{2}m^2_\rho\vec{\rho}_\mu\vec{\rho}^\mu.
\label{dlag}
\end{align}
in which $\psi$ represents the nucleon field, and $\sigma$, $\omega^\mu$, and $\vec{\rho}_\mu$ are the scalar, vector, and isovector-vector fields related to the respective mesons. Furthermore, we also have $F_{\mu\nu}=\partial_\nu\omega_\mu-\partial_\mu\omega_\nu$ and $\vec{B}_{\mu\nu}=\partial_\nu\vec{\rho}_\mu-\partial_\mu\vec{\rho}_\nu$. From Eq.~(\ref{dlag}) and the corresponding energy-momentum tensor~($T^{\mu\nu}$), it is possible to construct the energy density of the system and from that all, the remaining thermodynamical quantities of interest. Thus, from $\mathcal{E}=\left<T_{00}\right>$ one has
\begin{align} 
\mathcal{E} &=  \frac{m_{\sigma}^{2} \sigma^{2}}{2} +\frac{A\sigma^{3}}{3} +\frac{B\sigma^{4}}{4} 
-\frac{m_{\omega}^{2} \omega_{0}^{2}}{2} - \frac{Cg_{\omega}^4\omega_{0}^4}{4}
- \frac{m_{\rho}^{2} \bar{\rho}_{0(3)}^{2}}{2} 
\nonumber\\
&+ g_{\omega} \omega_{0} \rho_B + \frac{g_{\rho}}{2} 
\bar{\rho}_{0(3)} \rho_{3}   -\frac{1}{2} \alpha'_3 g_{\omega}^{2} g_{\rho}^{2} \omega_{0}^{2} 
\bar{\rho}_{0(3)}^{2}
+ \mathcal{E}_{\mathrm{kin}}^{p}+\mathcal{E}_{\mathrm{kin}}^{n},
\label{eden}
\end{align}
where $\sigma$, $\omega_0$, and $\bar{\rho}_{0(3)}$ are the mean-field quantities related to the mesons fields, obtained from the solution of the field equations, $\rho_3=(2Y_p-1)\rho_B$, $Y_p$ is the proton fraction, and $\rho_B$ is the baryonic density. 
The interested reader can obtain all the details of the EOS calculation in many papers in the literature, as for instance, in~\cite{rev3,dutra2014,debora-universe}
As long as the SRC correlations are included in the system, the protons and neutrons kinetic parts of the energy density become
\begin{eqnarray} 
\mathcal{E}_{\text {kin }}^{n,p} &=& \frac{\gamma \Delta_{n,p}}{2\pi^2} \int_0^{{k_{F\,{n,p}}}} 
k^2dk({k^{2}+M^{* 2}})^{1/2}
\nonumber\\
&+& \frac{\gamma C_{n,p}}{2\pi^2} \int_{k_{F\,{n,p}}}^{\phi_{n,p} {k_{F\,{n,p}}}} 
\frac{{k_F}_{n,p}^4}{k^2}\, dk({k^{2}+M^{* 2}})^{1/2},\nonumber \\
\label{ekin}
\end{eqnarray}
with $M^*= M_N-g_\sigma\sigma$, and the Fermi momenta related to neutrons/protons denoted by ${k_F}_{n,p}$. This particular change is due to the new momentum distribution function induced by the SRC, namely,
\begin{eqnarray}
n_{n,p}(k) = \left\{ 
\begin{array}{ll}
\Delta_{n,p}, & 0<k<k_{F\,{n,p}}
\\ \\
\dfrac{C_{n,p}\,k_{F\,{n,p}}^4}{k^4}, & k_{F\,{n,p}}<k<\phi_{n,p} k_{F\,{n,p}}.
\end{array} 
\right.
\label{eqhtm}
\end{eqnarray}
This particular expression phenomenologically represents the depletion below the nucleon Fermi surface, and the momentum tail structure above it, both effects being a direct consequence of the nuclear force and experimentally verified (see Ref.~\cite{ppnp-src}, for instance, for a review of this subject. In addition to modifying the kinetic term of the energy density, such new distribution, defined at $T=0$, also affects the scalar densities, the chemical potentials of protons and neutrons, and the pressure of the system. The explicit expressions of these quantities can be seen in Ref.~\cite{lucas}. 
\\
In principle, the three parameters $\Delta_{n,p}$, $C_{n,p}$, and $\phi_{n,p}$ are independent quantities. However, the normalization condition, given by
\begin{equation}
\frac{1}{\pi^2} \int_0^{\infty}dk\,k^2\,n_{n,p}(k) = \frac{({k_{F\,{n,p}}})^3}{3\pi^2},
\label{norm}
\end{equation}
can be used to write, for instance, $\Delta_{n,p}$ as a function of $C_{n,p}$ and $\phi_{n,p}$. Furthermore, according to previous calculations by using self-consistent Green's function~\cite{green}, and Brueckner-Hartree-Fock~\cite{bhf} theories, the depletion $\Delta_{n,p}$ is almost a linear function of the isospin parameter $1-2Y_p$. This information along with the normalization condition is used to find
\begin{align}
\Delta_{n,p} &=1 - 3C_{n,p}(1-1/\phi_{n,p}),
\\
C_{n,p}&=C_0[1 \pm C_1(1-2Y_p)],
\\
\phi_{n,p}&=\phi_0[1 \pm \phi_1(1-2Y_p)]
\end{align}
with $(+)$ for neutrons and $(-)$ for protons. Concerning the values of the constants $C_{0,1}$, $\phi_{0,1}$, we use $C_0=0.161$, $C_1=-0.25$, $\phi_0 = 2.38$ and $\phi_1=-0.56$, respectively~\cite{cai}. These numbers ensure a fraction of $28\%$ of nucleons in the momentum tail part of the nucleon distribution in symmetric nuclear matter, and a fraction of $1.5\%$ in pure neutron matter~\cite{cai}.

In this work we use the parametrizations FSU2R and IUFSU with SRC included because these parameterizations have been shown to produce reliable results when applied to investigate subsaturation properties and also very high density matter, as the one present in the interior of neutron stars~\cite{PhysRevC.93.025806,PhysRevC.94.049901,PhysRevC.99.045202,FSU2R}.  Additionally, we use the NL3 with SRC, which neither satisfies high-density constraints in the pressure~\cite{danielewicz2002}, nor other nuclear bulk properties~\cite{dutra2014}. However, it is only considered here for the sake of comparison to better assess the role of the symmetry energy.
Their respective coupling constants, with and without this phenomenology implemented, are given in Table~\ref{tabconst}.
\begin{table*}[!htb]
\centering
\setlength{\tabcolsep}{4pt}
\caption{Coupling constants of the FSU2R, IUFSU and NL3 parametrizations with and without SRC included. The nucleon rest mass is $M_N=939$~MeV.
}
\begin{tabular}{lrrrrrr}
\hline\noalign{\smallskip}
coupling                       &  FSU2R   & FSU2R-SRC & IUFSU    & IUFSU-SRC  & NL3 & NL3-SRC\\
\noalign{\smallskip}\hline\noalign{\smallskip}
$g_\sigma$                     & $10.372$ & $10.517$  & $9.971$  & $10.132$ & $10.217$ & $10.496$ \\
$g_\omega$                     & $13.505$ & $12.365$  & $13.032$ & $11.867$ & $12.868$ & $12.013$ \\
$g_\rho$                       & $14.368$ & $15.599$  & $13.590$ & $15.551$ & $8.498$ & $14.849$ \\
$A/M_N$                          & $1.837$  & $2.913$  & $1.785$  & $2.956$  & $2.192X$ & $2.996$ \\
$B$                            & $-3.240$ & $-32.443$ & $0.488$  & $-29.880$ & $-28.885$ & $-45.697$ \\
$C$ ($\times 10^{-3}$)         & $4.000$  & $4.000$   & $5.000$  & $5.000$ & $0$ & $0$ \\
$\alpha_3'$ ($\times 10^{-2}$) & $9.000$  & $9.300$   & $9.200$  & $1.094$ & $0$ & $0.0192$ \\
\noalign{\smallskip}\hline
\label{tabconst}
\end{tabular}
\end{table*}

The parameters for the three models with SRC  were obtained by imposing that they reproduce the same bulk parameters of their versions without SRC, namely, saturation density ($\rho_0$), binding energy, effective mass, incompressibility, symmetry energy, and symmetry energy slope, all of them evaluated at $\rho_0$. The symmetry energy and its slope are defined as
\begin{equation}
S\left( \rho_B\right) = \left. \frac{1}{8} \frac{\partial ^2\left({\cal E}/\rho_B\right)}{\partial Y_p^2}  \right|_{Y_p=0.5}
\end{equation}
and
\begin{equation}
L(\rho_B) = 3 \rho_B \frac{\partial S }{\partial \rho_B},
\end{equation}
respectively. Their values at the saturation are named here as $J\equiv S(\rho_0)$ and $L_0\equiv L(\rho_0)$. It is worth mentioning that these original parametrizations were constructed in order to correctly describe some features of finite nuclei such as the neutron skin thickness, in the case of the IUFSU, for instance. In this initial work, we did not account for constraints from finite nuclei.

\begin{table}[!htb]
\centering
\caption{Bulk parameters of the FSU2R, IUFSU and NL3 parametrizations. They are identical with and without SRC.}
\begin{tabular}{lrrr}
\hline\noalign{\smallskip}
                        &  FSU2R &  IUFSU  & NL3 \\
\noalign{\smallskip}\hline\noalign{\smallskip}
    B/A (MeV)                       & $-16.3$   & $-16.4$     & $-16.3$  \\
    $K$ (MeV)                        & $238$    & $231.2$     &     $272$     \\
    $J$ (MeV)                       & $30.7$   & $31.3$      & $37.4$  \\
    $L_0$ (MeV)                        & $46.9$  & $47.2$      & $118.32$     \\
    $\rho_0$ (fm$^{-3}$)            & $0.150$  & $0.155$      & $0.148$    \\
    $M^\ast/M$                      & $0.59$   & $0.62$      & $0.60$     \\
\noalign{\smallskip}\hline
\label{tabbulk}
\end{tabular}
\end{table}

\subsection{Pasta phase}

In these calculations, as generally done when homogeneous matter is compared with the pasta phase, neutron-proton-electron (npe) matter is considered, i.e., not only nucleons, but also electrons that are responsible for a charge neutral matter, are taken into account.

As already mentioned, we use the simplest possible prescription, in which Gibbs conditions are enforced and the electron gas remains in the background. The two phases are generally referred as liquid (the denser one) and gas phase. In symmetric matter the gas phase at zero temperature is made purely of neutrons, but as the proton fraction decreases, protons also drip from nuclei and are incorporated in the gas phase.

The Gibbs conditions that have to be satisfied by the CPA method are given in~\cite{oldpasta}. They are:
\begin{equation}\label{eq:gibbs}
\begin{split}
        P^I = P^{II} \\
        \mu_p^I = \mu_p^{II} \\
        \mu_n^I = \mu_n^{II}
\end{split}
\end{equation} 
 The surface tension coefficient is a crucial quantity in this formalism and we have used a fitting to the Thomas-Fermi calculation as a function of the proton fraction calculated in~\cite{PRC85} for the IUFSU and for the NL3 models. At $T=0$, it reads:
\begin{equation}\label{sigpar} 
\begin{split}
\sigma (x,T=0) =\sigma_0& \exp{\left(-\sigma_1 x^{3/2}\right)} \big(1+a_1 x+ a_2 x^2\\
&+a_3 x^3 +a_4 x^4 + a_5 x^5 +a_6 x^6\big)~,
\end{split}
\end{equation}
where  $x=\left(  1-2 Y_p \right) ^2$  stands for  the squared relative neutron excess and  $\sigma_0$ is the surface coefficient for symmetric matter. The fitting parameters are given in Table~\ref{tabsig}.
\begin{table*}[!htb]
\begin{center}
\caption{Surface tension coefficient parameters fitted within the Thomas-Fermi approximation for the IUFSU and NL3 parametrizations. $\sigma_0$ is in MeV fm$^{-2}$.}
\label{tabsig}
\begin{tabular}{ccccccccc}
\hline\noalign{\smallskip}
Par. & $\sigma_0$ &  $\sigma_1$& $a_1$ & $a_2$ & $a_3$ & $a_4$ & $a_5$ & $a_6$ \\			
\hline\noalign{\smallskip}
IUFSU   & $1.16473$ & $-0.659167$   & $-2.25482$ &  $-5.64237$  & $37.8471$     & $-81.6617$    & $81.2696$     & $-31.0227$ \\
NL3     & $1.12307$ & $20.7779$     & $-5.84915$ & $138.839$    & $-1631.42$   & $8900.34$     & $-21592.3$      & $20858.6$ \\
\hline\noalign{\smallskip}
\end{tabular}
\end{center}
\end{table*}
 
 For the FSU2R, we used a fixed surface tension $\sigma_0=1.2$ MeV\,fm$^{-2}$, since no fitting is available in the literature. We are aware that this is a naive approximation, but we would like to obtain our results with two independent functionals to see whether the results are model dependent and we are certain that the results would only change quantitatively had we chosen more appropriate values. We have checked that for higher values of the surface interaction (around 1.4 MeV\,fm$^{-2}$), the pasta disappears at lower densities and even lower if SRC are included. For lower values (around $1.03$~MeV\,fm$^{-2}$), the opposite happens, i.e., the pasta survives up to higher densities and SRC effects also decrease the transition density.

Hence, neither the Coulomb, nor the surface energy have been chosen to depend on the SRC. However, the symmetry energy is a quantity that depends explicitly on the SRC and has indirect consequences on the appearance of the pasta phase. One can see the expression of the symmetry energy with SRC, for instance, in~\cite{cai}.

The geometries considered are droplets and bubbles (3D), rods and tubes (2D), and slabs (1D). In this work, fluctuations are not taken into account as done in~\cite{flut1,Pelicer:2021ils}, but an extension to take them into account is planned for a future work. Hence, to obtain the preferential ground state and its corresponding geometry at each density, the homogeneous phase is compared with the pasta phase for each internal structure. The lower one in free energy is the  preferential matter. 

\section{Results}

We next display some of the graphs for the IUFSU parameterization only because they are qualitatively similar to the ones obtained for the FSU2R (and also NL3, as we discuss next).  We have chosen to show results for symmetric matter ($Y_p = 0.5$), $Y_p=0.35$, usually considered in core-collapse supernova and $Y_p=0.1$, a quantity that relates to the amount of protons in neutron star matter. 

\begin{figure}[!htb]
\includegraphics[scale=0.65]{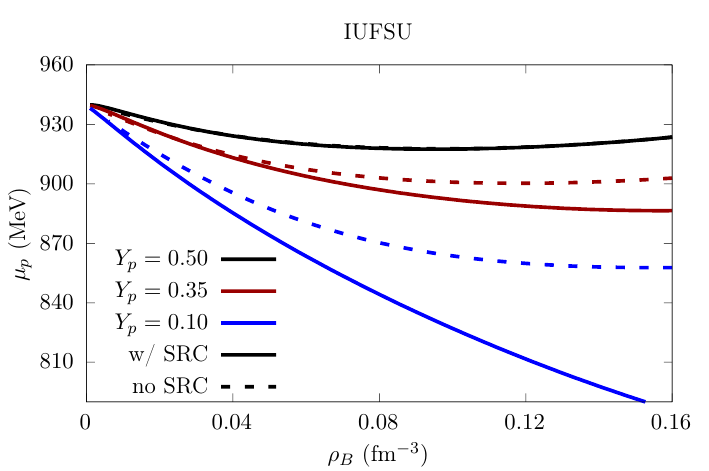}
\includegraphics[scale=0.65]{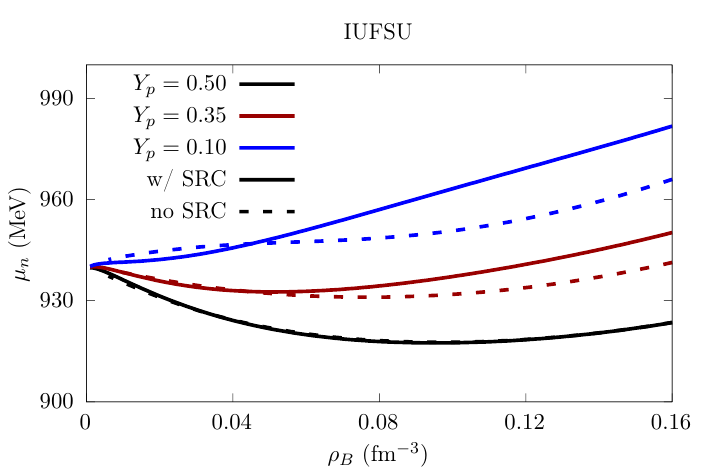}
\caption{Chemical potential of protons (top) and neutrons (bottom) for homogeneous matter with the IUFSU parametrization for different proton fractions, both without (dashed curves) and with (full curves) short range correlations.}
\label{fig:mus}
\end{figure}

 In Fig. \ref{fig:mus}, we compare the proton and neutron chemical potentials with and without SRC for different proton fractions. It is notorious that very asymmetric matter is more susceptible to changes than symmetric nuclear matter,  i.e., when one looks at the blue curves ($Y_p=0.1$) for both chemical potentials, the deviation when SRC are included is much larger than when one observes the black curves ($Y_p=0.5$).

\begin{figure}[!htb]
\includegraphics[scale=0.65]{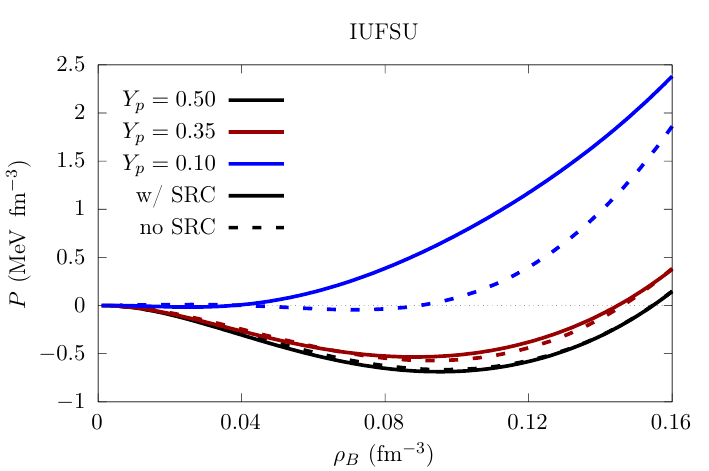}
\caption{Pressure of homogeneous baryon matter for different proton fractions with IUFSU without (dashed curves) and with (full curves) short range correlations for different proton fractions.}
\label{fig:press}
\end{figure}

As the main objective of this work is to investigate the influence of SRC on the pasta phase and the pressure is an important quantity in its construction, in Fig. \ref{fig:press} we also show how the pressure behaves as a function of the density. For asymmetric matter with SRC the pressure is less concave and starts increasing at smaller densities when compared to the original parametrization. A similar situation happens at the chemical potential: the neutron (proton) one, with SRC, becomes larger (smaller) around $0.05$~fm$^{-3}$, and becomes increasingly different from the original parametrization as density rises.   

\begin{figure}[!htb]
\includegraphics[scale=0.75]{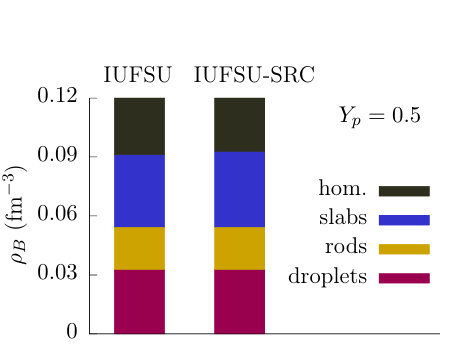}
\includegraphics[scale=0.75]{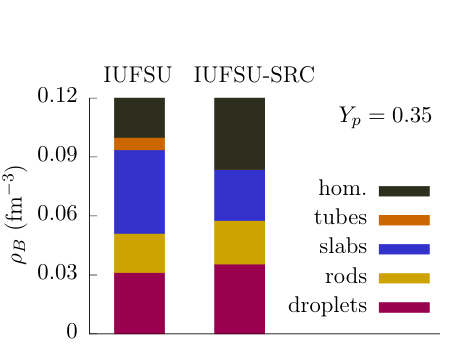}
\includegraphics[scale=0.75]{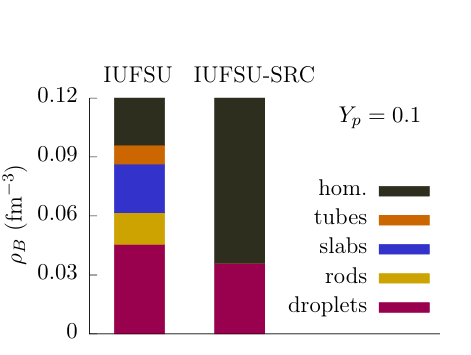}
\caption{Baryon densities where each pasta geometry is dominant for the IUFSU (left) and IUFSU-SRC (right) for proton fractions $Y_p=$0.5, 0.35 and 0.1 (left, center and right, respectively).}
\label{fig:pasta}
\end{figure}

\begin{figure}[!htb]
\includegraphics[scale=0.75]{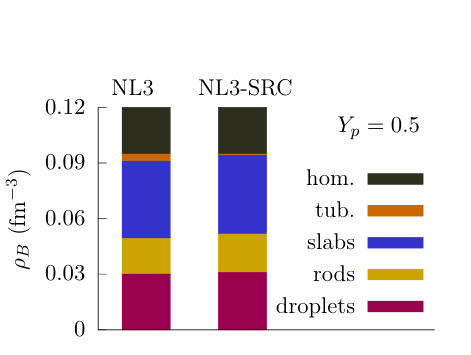}
\includegraphics[scale=0.75]{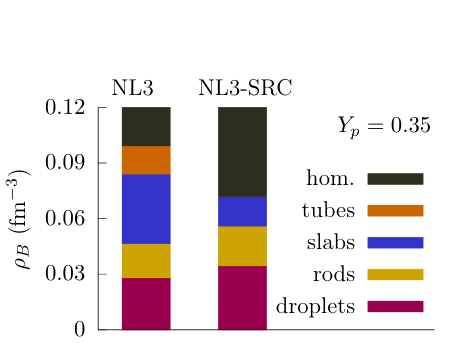}
\includegraphics[scale=0.75]{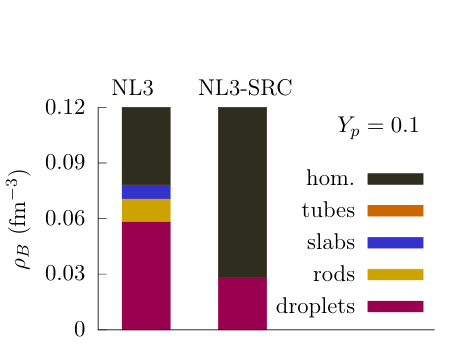}
\caption{Same as Fig.~\ref{fig:pasta} for the NL3 parametrization.}
\label{fig:pasta_nl3}
\end{figure}

\begin{figure}[!htb]
\includegraphics[scale=0.75]{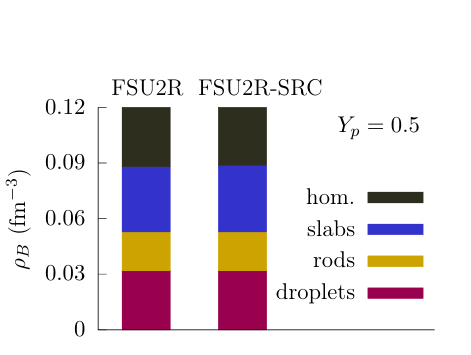}
\includegraphics[scale=0.75]{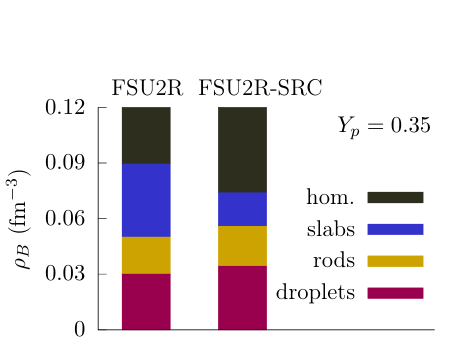}
\includegraphics[scale=0.75]{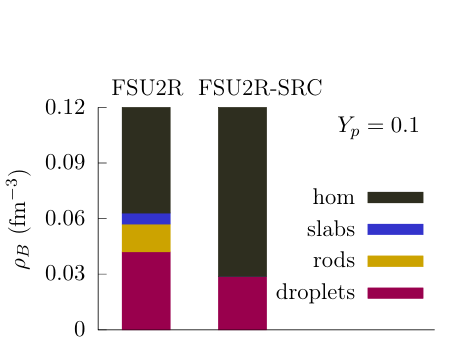}
\caption{Same as Fig.~\ref{fig:pasta} for the FSU2R parametrization.}
\label{fig:pasta_fsu2r}
\end{figure}

Once we have understood how the homogeneous matter is affected, we can obtain the pasta phase and check its structure using the Gibbs conditions \eqref{eq:gibbs}. In Figs.~\ref{fig:pasta},~\ref{fig:pasta_nl3} and~\ref{fig:pasta_fsu2r}, our results are shown for the IUFSU, NL3 and FSU2R models, respectively without (left) and with (right) SRC. As a consequence of the already observed differences in homogeneous matter, the larger the asymmetry, the stronger the effects of the SRC. 
For symmetric matter, the results are very similar, both as far as the size of the pasta phase as well as its internal structure. For $Y_p=0.35$, the similarities disappear: not only the pasta phase shrinks when SRC are included, but one of its internal structure, the tubes, vanishes. When we analyse $Y_p=0.1$, the most important result for the neutron star inner crust, the pasta phase diminishes even more and only droplets survive. 

\begin{figure*}[!htb]
    \centering
    \includegraphics[scale=0.7]{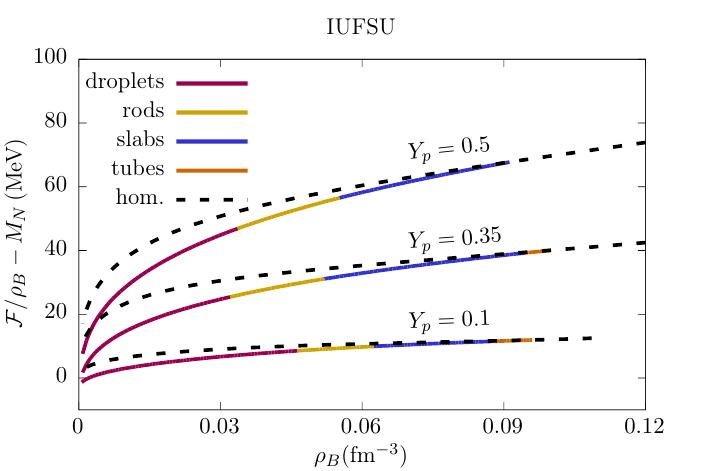}
    \includegraphics[scale=0.7]{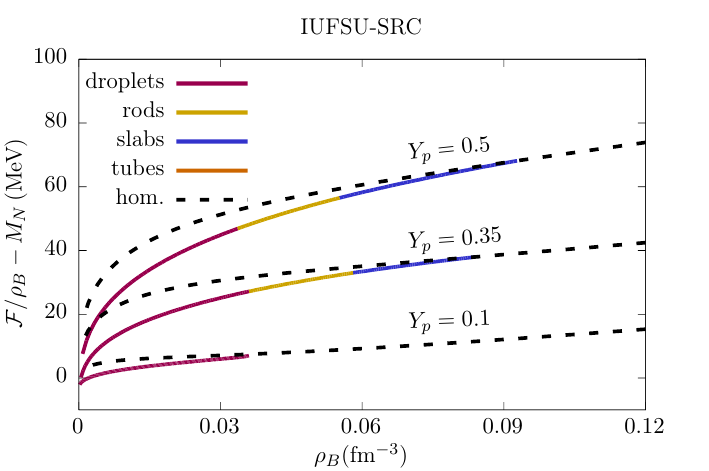}
    \caption{Free energy density per baryon of homogeneous npe matter (dashed curves) and pasta (full curves) with (top) and without (bottom) short range correlations. }
    \label{fig:pasta_energy}
\end{figure*}

For the sake of completeness, we also show the free energy density for different proton fractions without and with SRC in Fig. \ref{fig:pasta_energy}, from where the densities related to each pasta structure are obtained.

\begin{figure}[!htb]
    \centering
    \includegraphics[scale=0.75]{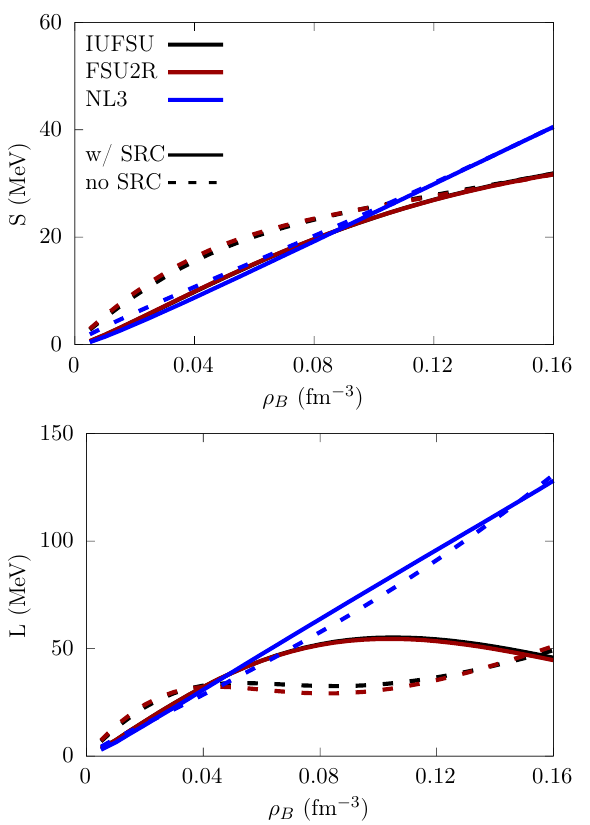}
    \caption{Symmetry energy and its slope for the IUFSU, FSU2R and NL3 parametrizations with (full line) and without (dashed line) SRC. }
    \label{fig:esym}
\end{figure}

Even though the bulk parameters are kept fixed at saturation point in the models with and without SRC, at different densities they can be completely different.  Thus, the disappearance of the pasta is 
related to an interplay involving the changes in the symmetry energy and its slope at low densities and to the change in the chemical potential, which is modified even in the case of identical bulk parameters, due to the modified momentum distribution given by Eq.~(\ref{eqhtm}). The symmetry energy and the slope are shown in Fig.~\ref{fig:esym}. In Ref.~\cite{PhysRevC.75.015801} it was shown that lower symmetry energy values at subsaturation densities decrease the transition to homogeneous matter, in agreement with our results. Nevertheless, the complete disappearance of the pasta at low proton fractions is quite an extreme case, and it cannot be solely due to the smaller symmetry energy, indicating that the change in chemical potential is also an important ingredient in the disappearance of the pasta. Higher order parameters, such as $K_{\rm sym}$, diminish due to the inclusion of SRC, but there is an insufficient correlation between the crust--core transition and $K_{\rm sym}$ to justify the complete disappearance of pasta~\cite{Balliet:2020nsh}.  
  At this point, a word of caution is necessary:
more robust calculations in which the surface tension can be self-consistently calculated, as explained in~\cite{PRC82,PRC85}, must be carried out when the SRC are included. The modifications with respect to the fitting obtained without SRC are certainly minor, but the quantitative results can be slightly modified. 

\section{Final remarks}

In this work, we investigate the effect of the short-range correlations on the pasta phase through a finite-range relativistic mean-field model. Very recently, the pasta phase was obtained with a point-coupling (zero-range) model for two different values of the slope (40 and 113 MeV) but similar symmetry energies at subsaturation densities
\cite{PhysRevC.103.055802}. It would be nice to understand if short-range correlations could be considered in that kind of model 
 as well. The SRC reduce the second derivative of the symmetry energy, but such a decrease does not seem to have an obvious correlation with the reduction of the crust--core transition density~\cite{Balliet:2020nsh}. 

Concerning our study, by analysing the present results in the face of the knowledge we have accumulated from previous works, we can guess two important outcomes for the inner crust of neutron stars:
1) as the temperature increases just to a few keV, probably no pasta phase will survive if SRC are considered in the mean field model, since temperature destroys the long-range ordering of the pasta~\cite{WATANABE2000455,PhysRevD.106.063020}, and if so 2) the calculation of transport properties~\cite{Yakovlev2016,Pelicer:2022ncw} can be greatly simplified, as the breaking of the clusters spherical symmetry that could lead to anisotropies in the collision frequencies would be greatly inhibited. Nevertheless, the SRC would also modify transport properties, adding yet more complexity to an already complicated formalism.
These features will be analyzed in future works.

\section*{ACKNOWLEDGMENTS}

This  work is a part of the project INCT-FNA Proc. No. 464898/2014-5. D.P.M. is  partially supported by Conselho Nacional de Desenvolvimento Cient\'ifico e Tecnol\'ogico (CNPq/Brazil) under grant  301155.2017-8  and  M.R.P. is supported  with a doctorate  scholarship by partly by Coordena\c c\~ao de Aperfei\c coamento de Pessoal de N\'ivel Superior (Capes/Brazil) and partly by CNPq. The work is also supported by CNPq under Grants No. 312410/2020-4 (O.L.) and No. 308528/2021-2 (M.D.).  O.L. and M.D. also acknowledge Funda\c{c}\~ao de Amparo \`a Pesquisa do Estado de S\~ao Paulo (FAPESP/Brazil) under Thematic Project 2017/05660-0 and Grant No. 2020/05238-9. O.~L. is also supported by FAPESP under Grant No. 2022/03575-3 (BPE). This study was financed in part by the Coordenação de Aperfeiçoamento de Pessoal de Nível Superior – Brasil (CAPES) – Finance Code 001 - Project number 88887.687718/2022-00 (M.D.).

\bibliographystyle{epja}
\bibliography{references}

\begin{thebibliography}{57}

\bibitem{PhysRevLett.50.2066}
D.G. Ravenhall, C.J. Pethick, J.R. Wilson, Phys. Rev. Lett. \textbf{50}, 2066
  (1983)

\bibitem{hashimoto84}
M.~Hashimoto, H.~Seki, M.~Yamada, Progress of Theoretical Physics \textbf{71},
  320 (1984)

\bibitem{Pons2013}
J.~Pons, D.~Vigan\'o, N.~Rea, Nature Physics pp. 431--434 (2013)

\bibitem{soft}
H.~Sotani, Monthly Notices of the Royal Astronomical Society: Letters
  \textbf{417}, L70 (2011)

\bibitem{Okamoto:2013tja}
M.~Okamoto, T.~Maruyama, K.~Yabana, T.~Tatsumi, Phys. Rev. C \textbf{88},
  025801 (2013), \texttt{1304.4318}

\bibitem{PhysRevC.83.035803}
M.D. Alloy, D.P. Menezes, Phys. Rev. C \textbf{83}, 035803 (2011)

\bibitem{PhysRevC.70.065806}
C.J. Horowitz, M.A. P\'erez-Garc\'{\i}a, J.~Carriere, D.K. Berry,
  J.~Piekarewicz, Phys. Rev. C \textbf{70}, 065806 (2004)

\bibitem{Lin:2020nxy}
Z.~Lin, M.E. Caplan, C.J. Horowitz, C.~Lunardini, Phys. Rev. C \textbf{102},
  045801 (2020), \texttt{2006.04963}

\bibitem{latetime}
S.W. Li, L.F. Roberts, J.F. Beacom, Phys. Rev. D \textbf{103}, 023016 (2021)

\bibitem{neutrino}
C.J. Horowitz, D.K. Berry, M.E. Caplan, T.~Fischer, Z.~Lin, W.G. Newton,
  E.~O'Connor, L.F. Roberts, \emph{Nuclear pasta and supernova neutrinos at
  late times} (2016), \texttt{https://arxiv.org/abs/1611.10226}

\bibitem{flut1}
C.C. Barros, D.P. Menezes, F.~Gulminelli, Phys. Rev. C \textbf{101}, 035211
  (2020)

\bibitem{Pelicer:2021ils}
M.R. Pelicer, D.P. Menezes, C.C. Barros, F.~Gulminelli, Phys. Rev. C
  \textbf{104}, L022801 (2021)

\bibitem{complex1}
F.J. Fattoyev, C.J. Horowitz, B.~Schuetrumpf, Phys. Rev. C \textbf{95}, 055804
  (2017)

\bibitem{complex2}
A.S. Schneider, M.E. Caplan, D.K. Berry, C.J. Horowitz, Phys. Rev. C
  \textbf{98}, 055801 (2018)

\bibitem{complex3}
B.~Schuetrumpf, G.~Mart\'{\i}nez-Pinedo, M.~Afibuzzaman, H.M. Aktulga, Phys.
  Rev. C \textbf{100}, 045806 (2019)

\bibitem{oldpasta}
S.S. Avancini, D.P. Menezes, M.D. Alloy, J.R. Marinelli, M.M.W. Moraes,
  C.~Provid\^encia, Phys. Rev. C \textbf{78}, 015802 (2008)

\bibitem{PRC82}
S.S. Avancini, S.~Chiacchiera, D.P. Menezes, , C.~Provid\^encia, Phys. Rev. C
  \textbf{82}, 055807 (2010)

\bibitem{PRC85}
S.S. Avancini, C.C. Barros, L.~Brito, S.~Chiacchiera, D.P. Menezes,
  C.~Provid\^encia, Phys. Rev. C \textbf{85}, 035806 (2012)

\bibitem{PhysRevC.105.025806}
W.G. Newton, S.~Cantu, S.~Wang, A.~Stinson, M.A. Kaltenborn, J.R. Stone, Phys.
  Rev. C \textbf{105}, 025806 (2022)

\bibitem{PhysRevD.106.063020}
C.J. Xia, T.~Maruyama, N.~Yasutake, T.~Tatsumi, Phys. Rev. D \textbf{106},
  063020 (2022)

\bibitem{PhysRevC.103.055810}
M.E. Caplan, C.R. Forsman, A.S. Schneider, Phys. Rev. C \textbf{103}, 055810
  (2021)

\bibitem{PhysRevC.102.015806}
F.~Ji, J.~Hu, S.~Bao, H.~Shen, Phys. Rev. C \textbf{102}, 015806 (2020)

\bibitem{sciencesrc1}
O.~Hen et~al., Science \textbf{346}, 614 (2014)

\bibitem{naturesrc2}
{CLAS Collaboration}, Nature \textbf{560}, 617 (2018)

\bibitem{naturescr3}
{CLAS Collaboration}, Nature \textbf{566}, 354 (2019)

\bibitem{naturesrc4}
A.~{Schmidt} et~al., Nature \textbf{578}, 540 (2020)

\bibitem{hen2017}
O.~Hen, G.A. Miller, E.~Piasetzky, L.B. Weinstein, Rev. Mod. Phys. \textbf{89},
  045002 (2017)

\bibitem{duer2019}
M.~{Duer} et~al., Physics Letters B \textbf{797}, 134792 (2019)

\bibitem{rev3}
B.A. Li, L.W. Chen, C.M. Ko, Physics Reports \textbf{464}, 113 (2008)

\bibitem{cai}
B.J. Cai, B.A. Li, Phys. Rev. C \textbf{93}, 014619 (2016)

\bibitem{baoanli21}
W.M. Guo, B.A. Li, G.C. Yong, Phys. Rev. C \textbf{104}, 034603 (2021)

\bibitem{baoanli22}
B.J. Cai, B.A. Li, Phys. Rev. C \textbf{105}, 064607 (2022)

\bibitem{baoanli-aop}
B.J. Cai, B.A. Li, Annals of Physics \textbf{444}, 169062 (2022)

\bibitem{lucas}
L.A. Souza, M.~Dutra, C.H. Lenzi, O.~Louren\ifmmode~\mbox{\c{c}}\else
  \c{c}\fi{}o, Phys. Rev. C \textbf{101}, 065202 (2020)

\bibitem{dmnosso1}
O.~Louren\ifmmode~\mbox{\c{c}}\else \c{c}\fi{}o, T.~Frederico, M.~Dutra, Phys.
  Rev. D \textbf{105}, 023008 (2022)

\bibitem{dmnosso2}
O.~Louren\ifmmode~\mbox{\c{c}}\else \c{c}\fi{}o, C.H. Lenzi, T.~Frederico,
  M.~Dutra, Phys. Rev. D \textbf{106}, 043010 (2022)

\bibitem{dmnosso3}
M.~Dutra, C.H. Lenzi, O.~Louren\c{c}o, Monthly Notices of the Royal
  Astronomical Society \textbf{517}, 4265 (2022)

\bibitem{science-src}
R.~Subedi et~al., Science \textbf{320}, 1476 (2008)

\bibitem{PhysRevC.72.015802}
T.~Maruyama, T.~Tatsumi, D.N. Voskresensky, T.~Tanigawa, S.~Chiba, Phys. Rev. C
  \textbf{72}, 015802 (2005)

\bibitem{IUFSU}
F.J. Fattoyev, C.J. Horowitz, J.~Piekarewicz, G.~Shen, Phys. Rev. C
  \textbf{82}, 055803 (2010)

\bibitem{FSU2R}
L.~Tolos, M.~Centelles, A.~Ramos, Publications of the Astronomical Society of
  Australia \textbf{34}, e065 (2017)

\bibitem{Lalazissis:1996rd}
G.A. Lalazissis, J.~Konig, P.~Ring, Phys. Rev. C \textbf{55}, 540 (1997),
  \texttt{nucl-th/9607039}

\bibitem{dutra2014}
M.~Dutra, O.~Louren\ifmmode~\mbox{\c{c}}\else \c{c}\fi{}o, S.S. Avancini, B.V.
  Carlson, A.~Delfino, D.P. Menezes, C.~Provid\^encia, S.~Typel, J.R. Stone,
  Phys. Rev. C \textbf{90}, 055203 (2014)

\bibitem{debora-universe}
D.P. Menezes, Universe \textbf{7} (2021)

\bibitem{ppnp-src}
J.~Arrington, D.~Higinbotham, G.~Rosner, M.~Sargsian, Progress in Particle and
  Nuclear Physics \textbf{67}, 898 (2012)

\bibitem{green}
A.~Rios, A.~Polls, W.H. Dickhoff, Phys. Rev. C \textbf{79}, 064308 (2009)

\bibitem{bhf}
P.~Yin, J.Y. Li, P.~Wang, W.~Zuo, Phys. Rev. C \textbf{87}, 014314 (2013)

\bibitem{PhysRevC.93.025806}
M.~Dutra, O.~Louren\ifmmode~\mbox{\c{c}}\else \c{c}\fi{}o, D.P. Menezes, Phys.
  Rev. C \textbf{93}, 025806 (2016)

\bibitem{PhysRevC.94.049901}
M.~Dutra, O.~Louren\ifmmode~\mbox{\c{c}}\else \c{c}\fi{}o, D.P. Menezes, Phys.
  Rev. C \textbf{94}, 049901 (2016)

\bibitem{PhysRevC.99.045202}
O.~Louren\ifmmode~\mbox{\c{c}}\else \c{c}\fi{}o, M.~Dutra, C.H. Lenzi, C.V.
  Flores, D.P. Menezes, Phys. Rev. C \textbf{99}, 045202 (2019)

\bibitem{danielewicz2002}
P.~Danielewicz, R.~Lacey, W.G. Lynch, Science \textbf{298}, 1592 (2002)

\bibitem{PhysRevC.75.015801}
K.~Oyamatsu, K.~Iida, Phys. Rev. C \textbf{75}, 015801 (2007)

\bibitem{Balliet:2020nsh}
L.E. Balliet, W.G. Newton, S.~Cantu, S.~Budimir, Astrophys. J. \textbf{918}, 79
  (2021), \texttt{2009.07696}

\bibitem{PhysRevC.103.055802}
F.~Ji, J.~Hu, H.~Shen, Phys. Rev. C \textbf{103}, 055802 (2021)

\bibitem{WATANABE2000455}
G.~Watanabe, K.~Iida, K.~Sato, Nuclear Physics A \textbf{676}, 455 (2000)

\bibitem{Yakovlev2016}
D.G. Yakovlev, Monthly Notices of the Royal Astronomical Society \textbf{453},
  581 (2015)

\bibitem{Pelicer:2022ncw}
M.R. Pelicer, M.~Antonelli, D.P. Menezes, F.~Gulminelli, Mon. Not. Roy. Astron.
  Soc. \textbf{521}, 743 (2023), \texttt{2212.11817}

\end{thebibliography}

\end{document}